\documentclass[prb,aps,showpacs,twocolumn,floatfix]{revtex4}
\usepackage{amsmath}
\usepackage{bbold}
\usepackage{graphicx}
\usepackage{xcolor}

\newcommand{\Mab}[4]{\left(\begin{array}{cc}
#1\ & #2 \\
#3\ & #4 \end{array} \right)}

\begin{document}

\title{Quartet currents in a biased three-terminal diffusive Josephson junction}
\date{\today}

\author{T. Jonckheere$^{1}$}
\author{J. Rech$^{1}$}
\author{C. Padurariu$^{2}$}
\author{L. Raymond$^{1}$}
\author{T. Martin$^{1}$}
\author{D. Feinberg$^{3,4}$}

\affiliation{$^1$ Aix Marseille Univ., Universit\'e de Toulon, CNRS, CPT, IphU, AMUtech, Marseille, France}
\affiliation{$^2$ Institut f\"ur Komplexe Quantensysteme and IQST, Universit\"at Ulm, Albert-Einstein-Allee 11, 89069 Ulm, Germany}
\affiliation{$^3$ Centre National de la Recherche Scientifique, Institut NEEL, F-38042 Grenoble Cedex 9, France}
\affiliation{$^4$ Universit\'e Grenoble-Alpes, Institut NEEL, F-38042 Grenoble Cedex 9, France}

\begin{abstract}
Biasing a three-terminal Josephson junction (TTJ) with symmetrical voltages $0,V,-V$ leads to new kinds of DC currents, namely quartet Josephson currents and phase-dependent multiple Andreev reflection (MAR) currents. 
We study these currents in a system where a normal diffusive metallic node $N$ is connected to three terminals $S_{0,1,2}$ by barriers of arbitrary transparency. We use the quantum circuit theory to calculate the current in each terminal, including decoherence.
 In addition to the stationary combination $\varphi_Q=\varphi_1+\varphi_2-2\varphi_0$ of the terminal phases $\varphi_i$, the bias voltage $V$ appears as a new and unusual control variable for a DC Josephson current. 
  A general feature is the sign changes of the current-phase characteristics, manifesting in minima of the quartet ``critical current". 
   Those sign changes can be triggered by the voltage, by the junction transparency or by decoherence. We
   study the possible separation of quartet currents from MAR currents in different regimes of parameters,
   including an "funnel" regime with very asymmetric couplings to $S_{0,1,2}$. In the regime of low transparency and asymmetric couplings, we provide an analytic perturbative expression
    for the currents which shows an excellent agreement with the full numerical results.
\end{abstract}

\pacs{
	73.23.-b,     
	73.63.Kv     
	74.45.+c    
}

\maketitle

\section {Introduction}
The understanding of subgap transport in transparent Superconducting/Normal Metal/Superconducting ($SNS$) Josephson junctions, either point contact-like, or made of a diffusive metallic region, has reached an advanced state on both experimental and theoretical levels. Multiple Andreev reflections (MARs) allow to describe dissipative quasiparticle motion below the superconducting gap $\Delta$, where transport is assisted by Cooper pairs.~\cite{MARs} 
Electrons and holes cross the structure, being Andreev-reflected at each $NS$ interface. Each crossing provides the energy $eV$ for electrons and holes, in such a way that steps in the $I(V)$ characteristics appear around values of the voltage $V\equiv 2\Delta/ne$ ($n$ positive integer). The most quantitative understanding of MAR transport has been obtained for point contact junctions, where transport is fully phase-coherent.~\cite{QPC} On the other hand, in diffusive $SNS$ junctions, another important parameter appears due to the dephasing of electrons and holes between successive Andreev reflections. 
This leads to the so-called Thouless energy $E_{Th}$ as another relevant energy scale.~\cite{Thouless} $E_{Th}$ is proportional to the inverse of the dwell time $\tau_d$ for a quasiparticle diffusing across the structure. In transparent $SNS$ junctions, shorter than the superconducting coherence length, one has $\Delta < E_{Th}$, while $E_{Th}<\Delta$ on the contrary in long diffusive junctions. MARs in the incoherent regime have been extensively studied~\cite{Bezuglyi2000} and reproducible junctions with very good quality are routinely fabricated. 
A minigap forms at an energy scale set by $E_{Th}$, which also sets the scale of the junction critical current, e.g. $eI_c \simeq G_NE_{Th}$ where $G_N$ is the junction normal conductance. 

Recently, multiterminal structures, made of three or more superconducting terminals contacted by a single normal region, have been considered theoretically, both at equilibrium~\cite{EPJB2015,Rech,Mi-Akhmerov,MelinMAR,Riwar,Meyer,Pillet} and under voltage bias.~\cite{Cuevas,Houzet,Freyn,JonckheerePRB2013,AkhmerovTTJ,KeliriDoucot} Experiments have addressed either three terminals but two being equipotential,~\cite{Giazotto} or three independent terminals i.e. biased with different voltages, the latter corresponding to the present study. Experiments involve junctions made of a diffusive metal,~\cite{Pfeffer} clean nanowires~\cite{Ronen,Khan} or 2D electron gas~\cite{Pankratova,Graziano1,Graziano2} and multichannel graphene.~\cite{Draelos,Harvard,Finkelstein21,Finkelstein22,Zhang} 
Transport in multiterminal junctions enables DC phase-coherent transport at nonzero but commensurate voltages.~\cite{Cuevas,Houzet,Freyn} Indeed, setting to zero both voltage and phase at, say, terminal $S_0$, and biasing terminals $S_1$ and $S_2$ at voltages $V_1$ and $V_2$, due to the Josephson relation $\frac{d\varphi_i}{dt}=\frac{2eV_i}{\hbar}$, two relevant variables can be chosen within the sets $(\varphi_1, \varphi_2)$ and $(V_1,V_2)$. 
For instance, fixing $V_1+V_2=0$ leaves $V_1-V_2=2V$ and $\varphi_1+\varphi_2$ as independent control variables. This holds for any commensurate combination $nV_1+mV_2=0$,  $n\varphi_1+m\varphi_2$ being the relevant stationary phase variable. 

As a result, a stationary current component can appear as a DC resonance, similar to an equilibrium Josephson current, but also controlled by the voltage V. For $V_1+V_2=0$ such a current involves two entangled pairs emitted from $S_0$ crossing simultaneously and transmitted, one in $S_1$, the other in $S_2$. Those processes have been dubbed as ``quartet currents'' since the elementary ``exchange currency'' between one terminal and the two others is made of two Cooper pairs instead of one. 
Of course, they coexist with the standard AC Josephson current flowing between any pair of terminals. 

Another striking signature of the voltage commensurability is the presence of phase-dependent "phase-MAR" currents, where quasiparticle transport is accompanied by quartets instead of single pairs like in the two-terminal case.~\cite{JonckheerePRB2013,Padurariu2,QuartetSQUID} Contrarily to two-terminal MARs flowing between any pair of the terminals biased at $(0,V,-V)$, phase-MARs involve all three terminals like quartets.

Such quartet currents, and higher order resonances, are reminiscent of the synchronization of AC Josephson currents in junction arrays, well documented some decades ago,~\cite{Likharev} and recently revisited  in a more general context.~\cite{AkhmerovMultipair} In these downmixing phenomena, leading to DC currents in biased arrays, the underlying mechanism can be understood by the classical nonlinear dynamics of an equivalent RSJ (resistively shunted junction) circuit, and it is strongly conditioned by the circuit environment, i.e. by the impedance of the junction plus the external circuit at the AC Josephson frequency. 
In contrast, the phenomena addressed in this work are intrinsically quantum and fully mesoscopic. Just as MARs, they cannot be described by classical or quantum phase dynamics in a Josephson junction array, using a given current-phase characteristics for each junction.~\cite{Likharev,AkhmerovMultipair} Quartet and phase-MAR currents are indeed another consequence of multiple Andreev reflections, allowed only in a multiterminal structure. They open new phase-coherent channels in a dissipative structure, that cannot be deduced in a phenomenological way from independent two-terminal Josephson currents. Discriminating classical downmixing from mesoscopic multi-pair resonances and separating quartets from phase-MARs remains an experimental challenge for the future, that could be tackled with a suitable SQUID geometry.~\cite{QuartetSQUID}

Experiments in TTJ's~\cite{Pfeffer,Ronen,Harvard} have revealed clear signatures of Josephson-like currents in a low-bias regime ($eV \ll \Delta$), beyond the regime of DC Josephson effects between any pair of terminals, i.e. for currents higher than the critical currents flowing in any pair of terminals. In this low-bias range, MAR currents are strongly hampered by a necessarily large number of Andreev reflections. On the contrary, quartets need minimally four Andreev reflections, which makes the quartet current visible against MARs. In fact, it has been shown theoretically that the quartet current does not decrease like MAR currents when $eV$ goes to $0$ but instead enters a regime where the current sign changes several times with voltage, due to nonadiabatic resonances.~\cite{Regis2,Regis3} In Refs. \onlinecite{Pfeffer,Ronen,Harvard}, robust transport anomalies were found as a function of the two applied voltages $V_{1,2}$, and were interpreted by three ``quartet'' modes corresponding to two Cooper pairs simultaneously crossing the structure. In Ref. \onlinecite{Pfeffer} the junctions are long and the Thouless energy is much smaller than the superconducting gap. The observation of strong anomalies in a regime $E_{Th} < V < \Delta$, where the voltage is too large to allow coherent motion of individual pairs, is especially striking. In Refs. \onlinecite{Ronen,Harvard}  instead, the system is in a coherent regime and MARs can be observed at higher voltages. 

A recent experiment used a four-terminal junction where two of them are grounded and enclose a loop pierced by a flux.~\cite{Harvard} The flux dependence of the quartet current displayed the expected oscillating behavior, but with two new features : the presence of the $\frac{hc}{4e}$ periodicity signaling the quartet mode, and a phase inversion of the oscillation. The latter was ascribed to resonances within the effective Andreev spectrum, due to the underlying running phase originating from the voltage bias.~\cite{Regis1,Regis2,Regis3} This sheds light on the microscopic origin of quartets, viewed as "self-induced Shapiro steps" in Ref. \onlinecite{Cuevas}. Theory has, until now, addressed the case of a few channels,~\cite{Regis1,Regis2,Regis3} describing the junction with a few-level quantum dot. MARs in a TTJ have also been described with the help of scattering theory, in the full $(V_1,V_2)$ plane.~\cite{AkhmerovTTJ} Yet a microscopic description is still lacking for a diffusive junction with a very large number of channels.

\begin{figure}[t]
\centering
\includegraphics[width=0.48\textwidth]{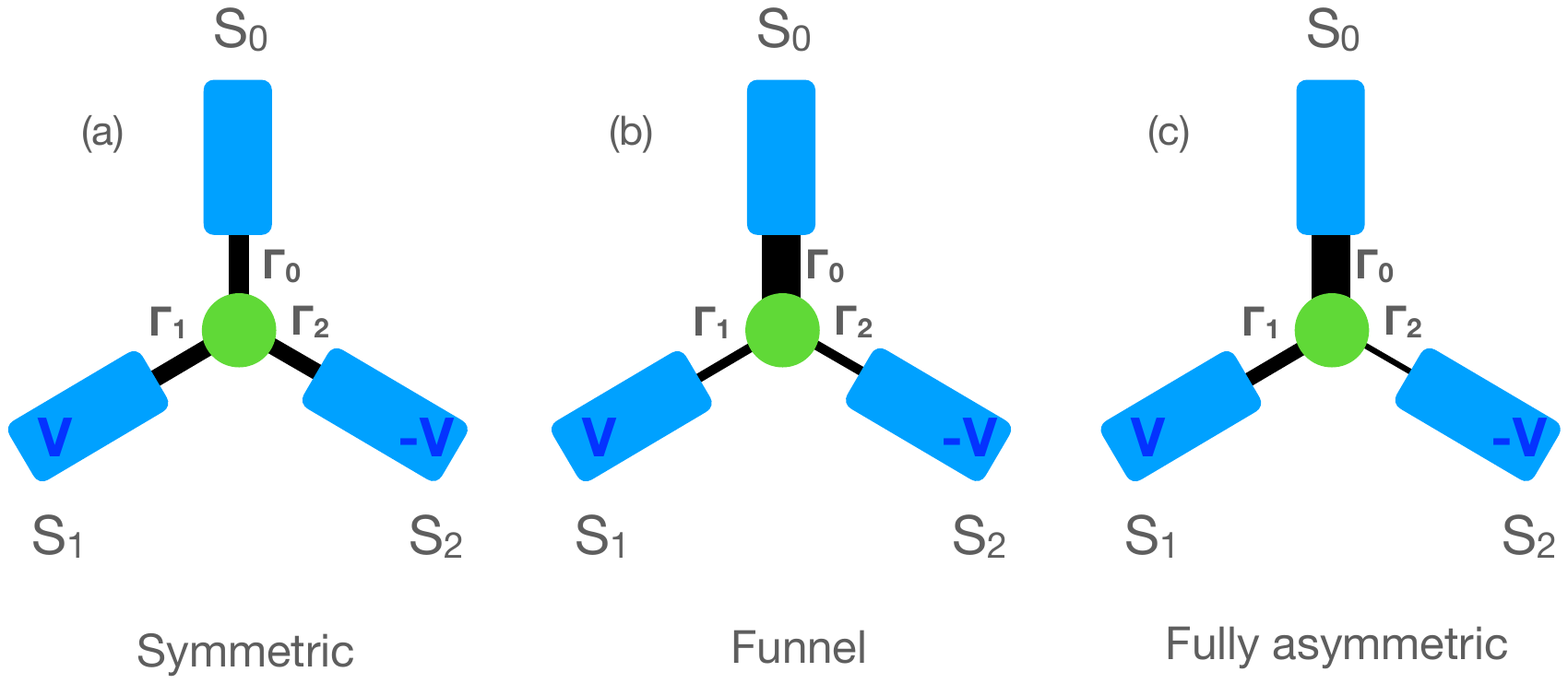}
\caption{Sketch of a three-terminal junction (TTJ), with superconductors $S_{0,1,2}$ at voltages $(0,V,-V)$ respectively, coupled to a node by couplings $\Gamma_{0,1,2}$ (see text). The thickness of the connection between the leads and the central node represents the strength of the corresponding coupling contant. (a) symmetric TTJ ($\Gamma_0=\Gamma_1=\Gamma_2$); (b) funnel TTJ ($\Gamma_0 \gg \Gamma_1,\Gamma_2$); (c) fully asymmetric TTJ ($\Gamma_0 \gg \Gamma_1 \gg \Gamma_2$).}
\label{fig:SchemaTTJ}
\end{figure}

Motivated by recent experiments where a sizable normal metal island (rather than a system of quantum dots) separates the superconducting leads, we consider here a short diffusive three-terminal junction (TTJ) (Fig. \ref{fig:SchemaTTJ}) described by quantum circuit theory~\cite{Yuli,Yulibook} to calculate quartet and MAR DC currents at any voltage. Dephasing can be introduced and may lead to a small Thouless energy, mimicking one important aspect of a long diffusive junction. The dependence of currents with voltage and quartet phase is studied in detail, both for a symmetric and a strongly asymmetric TTJ. This work follows previous ones where the spectral normal density of states NDOS was calculated, first at equilibrium,~\cite{Padurariu1} then at nonzero voltages~\cite{Padurariu2} at which BCS-like resonances appear at MAR voltages.  Moreover, the low-voltage behavior of the NDOS of a symmetric TTJ displays a phase-dependent pseudogap behavior,  allowing to make the link with an adiabatic regime when $V\rightarrow0$. 
In the present work, the spectral quartet current is shown to possess resonances at voltages coinciding with those of the NDOS, but with signs alternating between successive resonances. Integration on energy yields the low-voltage phase-coherent DC current, which in general departs strongly from an adiabatic behavior, with strong oscillations in the voltage dependence.

This motivates the study of an asymmetric "funnel" TTJ where the biased terminals are much more weakly coupled than the unbiased one (Fig. \ref{fig:SchemaTTJ}). In this situation, we show that it is possible to reach a quasi-adiabatic regime, where the low-voltage current does not exhibit sign changes anymore. The numerical study presented here is complemented by a perturbative expansion of the nonlinear circuit theory equations.

The other focus of the paper is the balance between quartet and MAR currents, important in view of obtaining unambiguous signatures of quartet current in experiments. As expected, even at low voltages, both have the same order of magnitude for a symmetric TTJ corresponding to a fully resonant cavity making the junction. On the contrary, in the funnel case, the phase-dependent MAR component becomes negligible at low voltages. Yet the phase-independent MAR remains strong if the junction formed by the two biased terminals (at $V, -V$) is symmetric. Only in a completely asymmetric TTJ, all MAR processes can be made smaller than quartets at low voltage. Finally, owing to the odd (for quartets) and even (for MARs) symmetries in phase, the DC current in all branches, which combines quartet and MAR contributions, yields a  phase shift in the current-phase relationship and therefore makes TTJs a new source of tunable "$\varphi$-junction".

The paper is structured as follows. The model and basic equations are presented in Section II. Section III discusses how the symmetry with respect to voltage inversion allows a practical definition of quartet and MAR currents for a junction with general asymmetry of the contacts. Section IV addresses a perturbative expansion when the contacts are strongly asymmetric, in the "funnel" case. Section V provides results on the spectral currents, that can be compared to the NDOS structure. Section VI describes the phase and voltage dependence of the quartet currents, for a symmetric and then for an asymmetric TTJ. Section VII discusses MAR currents and compares them to quartet currents. Section VIII provides a conclusion and perspectives.

\section {The model}

Let us consider  a three-terminal junction formed by connecting three superconductors with a normal diffusive metallic region (Fig. \ref{fig:SchemaTTJ}). The superconducting coherence length being larger than both the Fermi wavelength and the elastic mean-free-path, one can employ the quasiclassical equations of nonequilibrium superconductivity, which take the form of a diffusive equation for the quasiclassical Keldysh-Nambu Green's function:~\cite{Larkin}
\begin{equation}
\check{G} = \Mab{\hat{G}^R}{\hat{G}^K}{0}{\hat{G}^A}\ , \quad \check{G}^2=\check{1},
\label{eq:defG}
\end{equation}
where $\hat{G}^{R,A,K}$ correspond to the retarded, advanced and Keldysh components respectively.
In addition to Keldysh-Nambu space (to be denoted with a check hat, $\check{G}$), the quasiclassical Green's function generally depends on two times (or energies)  and on spatial coordinates, $\check{G}(\textbf{E}, \textbf{x})$. The diffusive equation for the corresponding Green's function is also known as the Usadel equation (see also Ref. \onlinecite{Yulibook}).
\begin{align}
\frac{\partial}{\partial \textbf{x}}\left({\cal D}(\textbf{x})\check{G}\frac{\partial}{\partial \textbf{x}}\check{G}\right)-i\left[\check{H},\check{G}\right]=0,\quad
\check{H} = \Mab{\hat{H}}{0}{0}{\hat{H}};\\ \hat{H}=E \hat{\sigma}_z+\frac{1}{2}\Delta(\textbf{x})(i\hat{\sigma}_y+\hat{\sigma}_x)+\frac{1}{2}\Delta^*(\textbf{x})(i\hat{\sigma}_y-\hat{\sigma}_x). \notag
\end{align}
Here the products should be understood as a matrix product in Nambu-Keldysh space along with a convolution in the double time (or energy) representation. The Pauli matrices in Nambu space are here denoted with a hat by $\vec{\hat{\sigma}}={\hat{\sigma}_x, \hat{\sigma}_y, \hat{\sigma}_z}$, and ${\cal D}(\textbf{x})$ denotes the diffusion coefficient.

The Usadel equation can be seen as a conservation of the quasiparticle current density, $\check{j}(\textbf{x})$.
\begin{align}
\label{usadelcurrent}
\frac{\partial}{\partial \textbf{x}}\check{j}(\textbf{x})+\frac{ie^2\nu}{\hbar}\left[\check{H},\check{G}\right]=0;\quad \check{j}=-\sigma(\textbf{x})\check{G}\frac{\partial}{\partial \textbf{x}}\check{G}.
\end{align}

Here, $\nu$ is the electronic density of states and $\sigma(\textbf{x})$ is the conductivity. The two quantities are related by $\sigma = e^2{\cal D}\nu$.

Here we employ a discretized version of the quasiclassical equations, the so-called quantum circuit theory.~\cite{Yuli,Yulibook}
The superconducting terminals $S_i$ are assumed to be homogeneous and described by Keldysh-Nambu matrices $\check{G}_i$, and they are connected by a discrete ``node'' described by the unknown matrix $\check{G}_c$. This discretization requires that the nodes dimensions are much smaller than the superconducting coherence length. At equilibrium, the terminal Green's functions are functions of one energy variable and are given by:
\begin{align}
\label{fgeq}
\hat{G}^R_i = \frac{1}{\xi}\Mab{\epsilon}{\Delta_i}{-\Delta_i^*}{-\epsilon}\ ;\quad \hat{G}^A_i = -\frac{1}{\xi^*}\Mab{\epsilon^*}{\Delta_i}{-\Delta_i^*}{-\epsilon^*}\ ,
\end{align}
where complex energies have been introduced as $\epsilon=E+i0^+$ and $\xi=\sqrt{\epsilon^2-|\Delta|^2}$. Here $\Delta_i=|\Delta|e^{i\varphi_i}$. The positive vanishing complex part of $\epsilon$ is essential in view of the branch cut of the square root function in the complex plane.

The advanced and retarded Green's functions are related by $\hat{G}_i^A = -\hat{\sigma}_z \left( \hat{G}_i^{R} \right)^{\dagger} \hat{\sigma}_z$ and the Keldysh Green's function $\hat{G}_i^K$ is obtained from:
\begin{equation} 
\hat{G}_i^K=h(E) \left(\hat{G}_i^R-\hat{G}_i^A\right), \quad h(E)= \tanh \left( \frac{\beta E}{2} \right)
\label{fgeq2}
\end{equation}

In the present work, terminals $S_{0,1,2}$ are respectively set to voltages $V_0=0, V_1=V,V_2=-V$, and the corresponding phases obey the Josephson relation $\varphi_i = \varphi_{0,i}+\frac{2eV_i t}{\hbar}$. Here the reference phase $\varphi_0$ is set to zero. While the Green's function of terminal $S_0$ is given by Eqs.~\eqref{fgeq}-\eqref{fgeq2}, those of terminals $S_{1,2}$ correspond to an out-of-equilibrium situation and must be described by a two-time Green's function $\check{G}_j (t_1,t_2)$. The latter is connected to its equilibrium counterpart by the following gauge transformation:
\begin{equation}
\label{gaugetransfo}
\check{G}_i (t_1,t_2) = e^{i (\hat{\sigma}_z \otimes \mathbb{1}) eV_i t_1} \check{G}_i(t_1 - t_2) e^{-i (\hat{\sigma}_z \otimes \mathbb{1}) eV_i t_2}
\end{equation}
where $\hat{\sigma}_z$ acts in Nambu space while $\mathbb{1}$ acts in Keldysh space, and $\check{G}_i(t_1 - t_2)$ corresponds to the equilibrium Green's function given by Eqs.~\eqref{eq:defG} and \eqref{fgeq}-\eqref{fgeq2} (upon Fourier transforming back from energy  to time variable),

Due to the time-dependence in the Hamiltonian, and periodicity with a single Josephson frequency $\omega_0=\frac{2eV}{\hbar}$, one can use double-time Fourier transform and represent the Green's functions in terms of a single energy variable (within the interval $[-\frac{\omega_0}{2},\frac{\omega_0}{2}]$) along with two harmonic indices, $\check{G}(E_1 = E + n \omega_0 , E_2 = E + m \omega_0) \to \check{G}(E,n,m)$~\cite{Padurariu2}. Notice that this definition is redundant, as $\check{G}(E,n,m)=\check{G}(E-p\omega_0,n+p,m+p)$ for any integer $p$, such that calculations can be performed in the energy interval $[-\frac{\omega_0}{2},\frac{\omega_0}{2}]$.~\cite{Jonckheere2009} Convolution products in energy are then obtained as matrix products in the space of harmonics. An alternative representation with only one harmonic index has been used in Ref. \onlinecite{Bezuglyi2}, in this case the matrix products appear as a recursion relation in the harmonic indices. 

Let us now write the circuit theory equations. The node is separated from each terminal by a connector. Those express the conservation of the Keldysh matrix current, as a discretized version of Eq.~\eqref{usadelcurrent}. As shown by Nazarov,~\cite{Yuli} the (matrix) spectral current flowing between terminal $S_i$ and the node takes the form:
\begin{align}
\check{I}_{ic} &= \Mab{\hat{I}_{ic}^R}{\hat{I}_{ic}^K}{0}{\hat{I}_{ic}^A} \\
&=\frac{2 e^2}{\pi\hbar}\sum_n \frac{ T^{(i)}_n [\check{G}_i,\check{G}_c]}{4+T^{(i)}_n(\{\check{G}_i,\check{G}_c\}-2)}
\label{eq:currentformula}
\end{align}
where the $T^{(i)}_n$ are the transmission coefficients of channel $n$ across each connector. Note that since $\check{G}^2=\check{1}$, both matrices $\check{G}_i$ and $\check{G}_c$ commute with $\{\check{G}_i,\check{G}_c\}$. This explains the notation using the fraction for matrix inversion in Eq.~\eqref{eq:currentformula}. To simplify, one can formally take $N$ equivalent channels and define $T^{(i)}_n=T\Gamma_i/N$, where $T$ is a scaling transparency factor for all contacts, and the factors $\Gamma_i$ are used to describe contact asymmetries. 

The physical DC current  $I_i$  in terminal $i$ (index $c$ is omitted), as measured in an experiment, is then given by the Nambu trace of the Keldysh component of the matrix current $\hat{I}_{ic}^K$ as
\begin{equation}
I_i = \int_{-\omega_0/2}^{\omega_0/2} dE \sum_n \text{Tr}_N \left[  \hat{\sigma}_z \hat{I}_{ic}^K (E,n,n)  \right] .
\label{eq:physicalI}
\end{equation}
 
The circuit theory formulation allows to take into account the dephasing which takes place in a normal metal diffusive conductor. In the single node model the only source of coherence loss is due to the difference in wave-vectors of electrons and holes at the same energy. The loss of coherence is taken into account by connecting a fictitious terminal to the node. The current to this terminal is given by,
\begin{align}
\check{I}_{fc} = \frac{2 e^2}{\pi\hbar}\sum_i \frac{T\Gamma_i}{4} [\check{G}_f,\check{G}_c]\ ,\quad \check{G}_f = -i \frac{E\tau_d}{\hbar} \Mab{\hat{\sigma}_z}{0}{0}{\hat{\sigma}_z}.
\label{eq:Gf}
\end{align}
where $\hat{\sigma}_z$ is the Pauli matrix in Nambu space, and $\tau_d$ is the dwell time of quasiparticles in the node, including the connectors. Notice that no physical current flows towards this fictitious terminal. This can be checked by substituting the expression of $I_{fc}$ into Eq.\eqref{eq:physicalI}, which gives 0 thanks to the simple block-diagonal structure of $\check{G}_f$ with $\hat{\sigma}_z$ matrices on the diagonal.

The basic equation of circuit theory which determines $\check{G}_c$ stems from the law of current conservation:
\begin{align}
\sum_i \check{I}_{ic} + \check{I}_{fc}= 0\ .
\end{align}

The same equation can be re-written as a commutation relation between the Green's function of the central node and a matrix $\check{M}$ that depends both on the Green's functions of the terminals and that of the central node:
\begin{equation}
\label{currentconservation}
[\check{G}_c,\check{M}] = 0,
\end{equation}
\begin{equation}
\check{M} = \sum_{i=0,1,2}T\Gamma_i\left(\frac{\check{G}_i}{1+(T\Gamma_i/4)(\{\check{G}_i,\check{G}_c\}-2)}+\check{G}_f\right).
\label{eq:Mmatrix}
\end{equation}
The solution for the node Green's function can be written formally as:
\begin{equation}
\check{G}_c=\frac{\check{M}}{\sqrt{\check{M}^2}} ,
\label{eq:solGctunnel}
\end{equation}
where the denominator guarantees that $\check{G}_c$ obeys the normalisation condition $\check{G}_c^2 = \check{1}$. As $\check{M}$ depends on $\check{G}_c$, this is a highly non-linear equation.

The so-called "tunnel" approximation amounts to neglecting the term in $T \Gamma_i$ in the denominator of $\check{M}$, yielding:
\begin{equation}
\check{M} = \displaystyle{\sum_{i=\{0,1,2\}} }T\Gamma_i\left(\check{G}_i+\check{G}_f\right).
\end{equation}
This approximation requires that $T\ll 1$. 
One must stress that this approximation is much more than a simple perturbative calculation in the tunneling transmissions. Indeed, the circuit theory entails multiple scattering within the node. For instance, in the absence of decoherence, $\tau_d=0$, and for a symmetric junction with two terminals, the tunneling approximation turns out to be exact whatever the $T_n^{(i)}=T_n$.~\cite{VanevicBelzig2006} This is due to the fact that $\{G_c,G_i\}$ becomes independent of $i$ in this case and the denominator can be factorized in Eq.~\eqref{eq:Mmatrix}. This property does not hold for more than two terminals. 

In practice, the Green's functions and matrix currents are expressed in Nambu-Keldysh-harmonics space and thus have dimension $4(2N_m+1)$ where $N_m$ is the maximum number of harmonics of the Josephson frequency retained in the calculation. Satisfactory convergence is obtained by using $N_m=\frac{2\Delta}{e V}$ as the matrix entries decay quickly at large harmonics value. 
In the tunnel approximation, the $\check{M}$ matrix is independent from $\check{G}_c$ and can be computed from the terminals Green's functions. The solution for $\check{G}_c$ is given by Eq.\eqref{eq:solGctunnel}: $\check{G}_c$ is proportional to the $\check{M}$ matrix, but has to satisfy
the normalization condition $\check{G}_c^2 = \check{1}$. Numerically, this can solved at each energy by doing the following:
 $\check{M}$ is diagonalized by using the appropriate change of basis, its diagonal form is then normalized, so as to contain only $\pm 1$ eigenvalues, and then $\check{G}_c$ is obtained
 by returning to the original basis. In the general case, the nonlinear Eq.~\eqref{currentconservation} is solved by adding an iteration procedure. 
 Starting from an initial guess for $\check{G}_c$ (which can be for example the solution obtained in the tunnel approximation), the matrix $\check{M}$ is computed from
 Eq.\eqref{eq:Mmatrix}, and $\hat{G}_c$ is obtained from this $\check{M}$ matrix. From this new $\check{G}_c$, a new $\check{M}$ matrix can be computed, leading to a new
 $\check{G}_c$, etc. The iteration procedure is repeated until the change in $\check{G}_c$ from one step to the next is small enough.
Once the solution for $\check{G}_c$ is found, the current $I_i(\varphi_Q,V)$ in terminal $i$ is obtained from Eq.~\eqref{eq:physicalI}, where the summation is truncated, running from $n=-N_m$ to $+N_m$, where $N_m$ is the number of harmonics retained in the calculation. Alternatively, one can integrate the traced Keldysh component of the matrix current $\hat{I}_{ic}^K (E,0,0)$ over the interval $[-E_m,E_m]$, where $E_m$ is a large energy cutoff (corresponding here to $E_m = (N_m+1) \omega_0$).
In the following, unless explicitly stated, all the numerical results that we show have been obtained by solving the full non-linear equation using an iteration procedure.
 
\section{Quartet and MAR currents}

The three-terminal setup allows two distinct kinds of DC currents to flow within the structure. First, dissipative currents generalize ~\cite{Cuevas,Houzet,JonckheerePRB2013,AkhmerovTTJ} the MAR currents of a two-terminal junction. As a novel feature, those currents depend on the quartet phase, due to interfering scattering paths within the node.~\cite{JonckheerePRB2013,Padurariu2,QuartetSQUID} Second, the special condition $V_1=-V_2$ drives a DC Cooper pair component from terminal $S_0$ to terminals $S_{1,2}$, as a current of quartets (entangled pairs of Cooper pairs) $I_Q$. This current is dissipationless, like the excess current stabilized at a Shapiro step in a junction irradiated with microwaves.~\cite{Cuevas}

Thus one can write the Keldysh currents as:
\begin{equation}
\check{I}_{ic}=\check{I}_{Q,i}+\check{I}_{MAR,i}
\end{equation}
corresponding respectively to the quartets and MAR contributions. The MAR current can be further split as
\begin{equation}
\check{I}_{MAR,i}=\check{I}_{avMAR,i}+\check{I}_{phMAR,i}
\end{equation}
where $\check{I}_{avMAR,i}$ is the "ordinary" MAR current, obtained by averaging $\check{I}_{MAR}$ on the quartet phase, while $\check{I}_{phMAR,i}$ corresponds to the phase-dependent part. 

For a symmetric TTJ, such as $\Gamma_1=\Gamma_2$, the MAR current components flowing in terminals $S_1$ and $S_2$ are opposite, so that no MAR current flows through terminal $S_0$, whose current is only due to quartets. For this reason, we sometimes refer to $I_Q$ as the quartet current from $S_0$ in the symmetric case. In this case, the components $I_Q$ and $I_{MAR}$ are obtained by integrating over frequency [see Eq.~\eqref{eq:physicalI}] the spectral components defined as follows:

\begin{align}
\check{I}_{Q} & =\check{I}_{Q,0} = \check{I}_{0}=\frac{e^2}{h}T\Gamma_0[\check{G}_0,\check{G}_c], \\
\check{I}_{MAR} &= \check{I}_{MAR,1} = -\check{I}_{MAR,2} = \frac{e^2}{2h}T\Gamma_1\left[\check{G}_1-\check{G}_2,\check{G}_c \right]
\end{align}
where the products entail matrix products of the $\check{G}_{i,c}^{\alpha,\beta}(E,n,m)$ with $\alpha,\beta$ running on the Nambu indices $1$ (for electrons) and $2$ (for holes), and $n,m$ running as integers.

In the general case  of asymmetric contacts, one can no longer access a simple self-contained expression for the various components of the current. Instead, one may rely on the symmetries of the physical (energy integrated) current. Indeed, to separate the total current in its quartet and MAR components in the general case of asymmetric contacts, one may use the symmetries with respect to $V$ and $\varphi_Q$. Time-reversal symmetry actually imposes that the total current is odd under a change of sign of both voltage and phases. First, as any dissipative current, the MAR components are odd in voltage thus even in $\varphi_Q$. On the contrary, the quartet current is even in voltage and odd in $\varphi_Q$. Therefore one has for the DC currents:

\begin{eqnarray}
\label{symmetries1}
I_{Q}^{i}(\varphi_Q,-V)&=&I_{Q}^{i}(\varphi_Q,V)\\
I_{Q}^{i}(-\varphi_Q,V)&=&-I_{Q}^{i}(\varphi_Q,V)\\
I_{MAR}^{i}(\varphi_Q,-V)&=&-I_{MAR}^{i}(\varphi_Q,V)\\
I_{MAR}^{i}(-\varphi_Q,V)&=&I_{MAR}^{i}(\varphi_Q,V)
\label{symmetries4}
\end{eqnarray}
,   
This suggests to use symmetrization with respect to $V$ as a definition of MAR and quartet currents in each contact:

\begin{eqnarray}
I_Q^{i}(\varphi_Q,V)&=&\frac{1}{2}\big[I^{i}(\varphi_Q,V)+I^{i}(\varphi_Q,-V)\big]\\
I_{MAR}^{i}(\varphi_Q,V)&=&\frac{1}{2}\big[I^{i}(\varphi_Q,V)-I^{i}(\varphi_Q,-V)\big]
\end{eqnarray}

A word of caution is needed about those definitions, if used in the biased terminals. In fact, as found numerically and can be shown by symmetry considerations, in the symmetric TTJ case $\Gamma_1=\Gamma_2$, one obtains $I_Q^1=I_Q^2=-\frac{I_Q^0}{2}$. This corresponds to the intuition that a quartet current splits equally into $S_1$ and $S_2$. Conversely, in the general case $\Gamma_1\neq\Gamma_2$, one still finds $I_Q^0+I_Q^1+I_Q^2=0$ but $I_Q^1\neq I_Q^2$, except for low transparency, as shown by a perturbative expansion of the circuit theory equations (see next Section). This means that high order multiple Andreev reflections entangle different kind of processes, and the symmetrizing procedure is merely a convenient definition separating dissipative and nondissipative current components. Yet this definition is unambiguous in terminal $S_0$: for arbitrary couplings $\Gamma_i$, symmetrization in $V$ allows to isolate the quartet current.  Let us remark that this procedure is convenient in calculations, but less in experiments where extrinsic causes can perturb the symmetry between $V>0$ and $V<0$.

\section{Perturbative expansion} \label{sec:perturb}

Circuit theory is nonlinear, which manifests both in the normalization of the node Green's function and in its self-consistent definition through Eq.~\eqref{currentconservation}. It is nevertheless possible to find a perturbative solution when some of the three contacts are more weakly coupled than others. This is especially useful in the case where the weaker couplings are with the biased contacts, that we call a "funnel" TTJ (Fig. \ref{fig:SchemaTTJ}b).

Let us consider a TTJ with couplings $\Gamma_i$ between the node and the three terminals at voltages $0,V,-V$. The contact transparencies are $T_i=T\Gamma_i$ where $T$ denotes the overall transparency of the TTJ. When $T\ll1$, which is assumed here, the equations for the node Green's function $\check{G}_c$ (in Nambu-Keldysh-harmonics space) and the currents take the form:

\begin{eqnarray}
\label{eqsGc}
\check{M}&=&T\sum_{i=0,1,2}\Gamma_i(\check{G}_i+\check{G}_f)\\
\left[\check{G}_c,\check{M}\right]&=&0\\
(\check{G}_c)^2&=&1\\
\check{I}_i&=&\frac{e^2}{h}T \Gamma_i\left[\check{G}_i,\check{G}_c\right]
\label{eq:matcurrGc}
\end{eqnarray}
This leaves the couplings $\Gamma_i$ and the decoherence parameter $\tau_d$ (see Eq. \eqref{eq:Gf}) as the only parameters of the problem. We first  take $\check{G}_f=0$ (no decoherence).

Since we are focusing on the case of a funnel TTJ, let us define $\Gamma_0=\gamma_0$ ($=1$ by definition if unspecified),  $\Gamma_{1}=\varepsilon \gamma_{1}$ and 
$\Gamma_{2}=\varepsilon \gamma_{2}$ (with $\gamma_{1}$ and $\gamma_2 \sim \gamma_0$) where $\varepsilon\ll1$ is an expansion parameter. We know a priori that to obtain the quartet current we need at least a ``trajectory" (in terms of products of Green's functions) passing twice through terminal $0$ and once through both terminals $1,2$, thus second order in $\varepsilon$ is required. Let us write:

\begin{equation}
\check{G}_c=\check{G}_0+\varepsilon \check{X}+\varepsilon^2 \check{Y}
\end{equation}
and solve at order 2 for Eqs. (1-3). Defining $\check{G}_{12}=\gamma_1\check{G}_1+\gamma_2\check{G}_2$, this yields:

\begin{eqnarray}
\label{eqsXY}
\{\check{X},\check{G}_0\}&=&0\\
\left[\check{X},\check{G}_0\right]&=&\left[\check{G}_{12},\check{G}_0\right]\\
\{\check{Y},\check{G}_0\}&=&-\check{X}^2\\
\left[\check{Y},\check{G}_0\right]&=&\left[\check{G}_{12},\check{X}\right]
\end{eqnarray}
Owing to the constraint $\check{G}_0^2=1$, a solution for Eq.~\eqref{eqsXY} is: 

\begin{equation}
\check{X}=\check{Z}-\check{G}_0\check{Z}\check{G}_0
\end{equation}
where $\check{Z}$ is an arbitrary matrix. Inserting in the other equations and solving finally gives:

\begin{eqnarray}
\check{Z}&=&\frac{1}{2}\check{G}_{12}\\
\check{Y}&=&-\frac{1}{2}\big[\check{Z}\check{G}_0\check{Z}+\{\check{Z}^2,\check{G}_0\}-3\check{G}_0\check{Z}\check{G}_0\check{Z}\check{G}_0\big]
\end{eqnarray}

The current in terminal $0$ is obtained as an expansion in $\varepsilon$, starting at first order:
\begin{eqnarray}
\check{I}_0&=&\varepsilon \check{I}_0^{(1)}+\varepsilon^2 \check{I}_0^{(2)}\\
\check{I}_0^{(1)}&=&\frac{e^2}{h}T\left[\check{G}_0,\check{G}_{12}\right]\\
\check{I}_0^{(2)}&=&\frac{e^2}{2h}T^2\left[\check{G}_{12},\check{G}_0\check{G}_{12}\check{G}_0\right]
\end{eqnarray}

The first order term contains components involving terminal $0$ and one of terminals $(1,2)$, i.e. quasiparticle currents like two-terminal MARs. One verifies that the quartet current is contained in the second-order terms that involve all three terminals. This yields a simple expansion for the second order (quartet) current component:
\begin{eqnarray}
\label{eqCurPerturb}
\nonumber
\check{I}_0^{(2)}&=&\frac{e^2}{2h}T^2\big(\Gamma_1\Gamma_2)^2
(\check{G}_1\check{G}_0\check{G}_2\check{G}_0+\check{G}_2\check{G}_0\check{G}_1\check{G}_0\\
&-&\check{G}_0\check{G}_1\check{G}_0\check{G}_2-\check{G}_0\check{G}_2\check{G}_0\check{G}_1\big)
\end{eqnarray}

Let us show that, in a second order expansion, the quartet current components in $S_1$ and $S_2$, obtained by symmetrization in $V$, are equal. One indeed shows easily that they are given by:

\begin{eqnarray}
\label{eqCurCompPerturb}
\nonumber
\check{I}_{1}^{(2)}&=&\frac{e^2}{2h}T^2\big(\Gamma_1\Gamma_2)^2
\big(\check{G}_0\check{G}_2\check{G}_0\check{G}_1-\check{G}_1\check{G}_0\check{G}_2\check{G}_0\big)\\
\check{I}_{2}^{(2)}&=&\frac{e^2}{2h}T^2\big(\Gamma_1\Gamma_2)^2
(\check{G}_0\check{G}_1\check{G}_0\check{G}_2-\check{G}_2\check{G}_0\check{G}_1\check{G}_0\big)
\end{eqnarray}

\begin{figure}[t]
\centering
\includegraphics[width=0.45\textwidth]{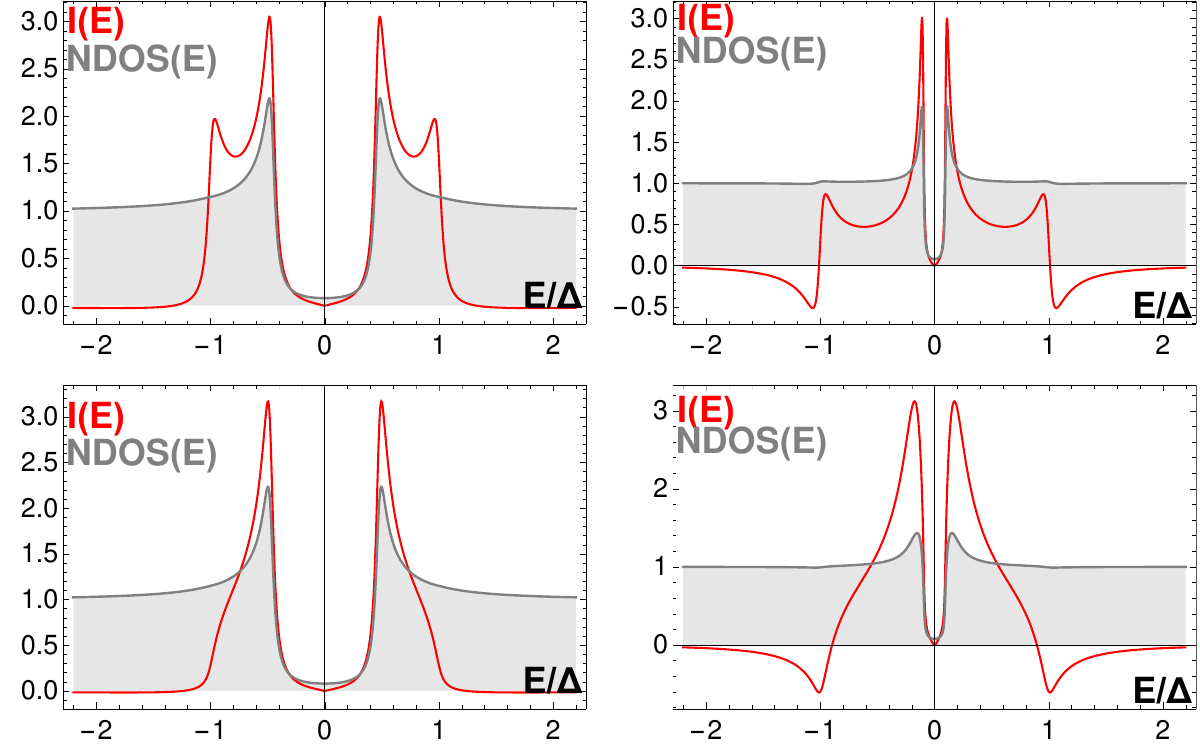}
\caption{Spectral current (red) and NDOS (grey) for a two-terminal junction at equilibrium. Scales are arbitrary. The spectral current peaks at the gap and minigap edges, especially at low transparency; (top, left) $T=0.1,\tau_d=0.025$; (top, right) $T=0.1,\tau_d=2$; (bottom, left) $T=0.8,\tau_d=0.025$; (bottom, right) $T=0.8,\tau_d=2$.}
\label{fig:SpecCurJJ}
\end{figure}

Replacing $V$ by $-V$ amounts to exchanging $\check{G}_1$ and $\check{G}_2$, as well as the phases $\varphi_1$, $\varphi_2$. Since the DC current is a function of $\varphi_1+\varphi_2$ only, one sees that it also amounts to exchanging $\check{I}_{1}^{(2)} $ and 
$\check{I}_{2}^{(2)}$, once integration on energy is performed. Therefore, symmetrization in $V$ yields the same quartet component in $S_1$ and $S_2$, equal to half of that in $S_0$ given by Eq.~\eqref{eqCurPerturb}. Pushing the above expansion to higher order ($3$ and $4$), one can verify that this property does not hold anymore if $\Gamma_1\neq \Gamma_2$. This shows that in the general case, if one is interested in the current in the biased terminals, symmetrization in $V$ allows to separate a "quartet" from a "MAR" DC component only for a symmetric junction. Contrarily, the symmetrizing procedure can always be used in terminal 0. Notice that antisymmetrization in $\varphi_Q$ is equivalent to symmetrization in $V$, owing to Eqs.~\eqref{symmetries1}-\eqref{symmetries4}. 

Taking into account decoherence, if $\check{G}_f$ is nonzero, the zeroth order in $\varepsilon$ for the node Green's function is obtained by solving:
\begin{eqnarray}
\left[\check{G}_c^{(0)},\check{G}_0+\check{G}_f\right]&=&0\\
\big(\check{G}_c^{(0)}\big)^2&=&1
\end{eqnarray}

This yields the same formal solution as in the coherent case, replacing $\check{G}_0$ by $\check{G}_c^{(0)}$:
\begin{eqnarray}
\check{G}_c&=&\check{G}_c^{(0)}+\varepsilon \check{X}+\varepsilon^2 \check{Y}\\
\check{Z}&=&\frac{1}{2}\check{G}_{12}\\
\check{X}&=&\check{Z}-\check{G}_c^{(0)}\check{Z}\check{G}_c^{(0)}\\
\nonumber
\check{Y}&=&-\frac{1}{2}\big[\check{Z}\check{G}_c^{(0)}\check{Z}+\{\check{Z}^2,\check{G}_c^{(0)}\}-3\check{G}_c^{(0)}\check{Z}\check{G}_c^{(0)}\check{Z}\check{G}_c^{(0)}\big]\\
\end{eqnarray}
and the current in terminal $0$:
\begin{equation}
\check{I}_0=\frac{e^2}{h}T\left[\check{G}_0,\check{G}_c^{(0)}\right]
\end{equation}
which does not simplify like in Eq.~\eqref{eqCurPerturb}.

This perturbative calculation can be generalized to  arbitrary transparencies.

\section{Spectral quartet currents}
In this Section, we discuss the spectral structure of the current components in a symmetric TTJ ($\Gamma_1=\Gamma_2$) biased at ($0, V,-V$), calculated in the full energy interval, in units of $\frac{e^2}{h}$. They are given by ${\cal I}_{ic}( E)=\Re (\check{I}_{ic}^{24}(E,0,0)-\check{I}_{ic}^{13}(E,0,0))$, and they are compared to the spectral density (NDOS) ${\cal N}_{c}( E)=\Re (\check{G}_{c}^{11}(E,0,0)-\check{G}_{c}^{22}(E,0,0))$, as calculated in Ref. \onlinecite{Padurariu2}. 

As a benchmark, Fig. \ref{fig:SpecCurJJ} shows the spectral current and the NDOS for a two-terminal junction {\it at equilibrium}, at fixed phase, varying the decoherence ($\tau_d$) and the transparency $T$. In both spectra, one sees the minigap, which is reduced by decoherence. The spectral current shows a sharp peak at the minigap edges, plus a marked peak around the gap in the tunnel case ($T \ll 1$, top panels of Fig. \ref{fig:SpecCurJJ}). The structures are broadened at high transparency. All take locally the form of a renormalized BCS NDOS. 

\begin{figure}[t]
\centering
\includegraphics[width=0.4\textwidth]{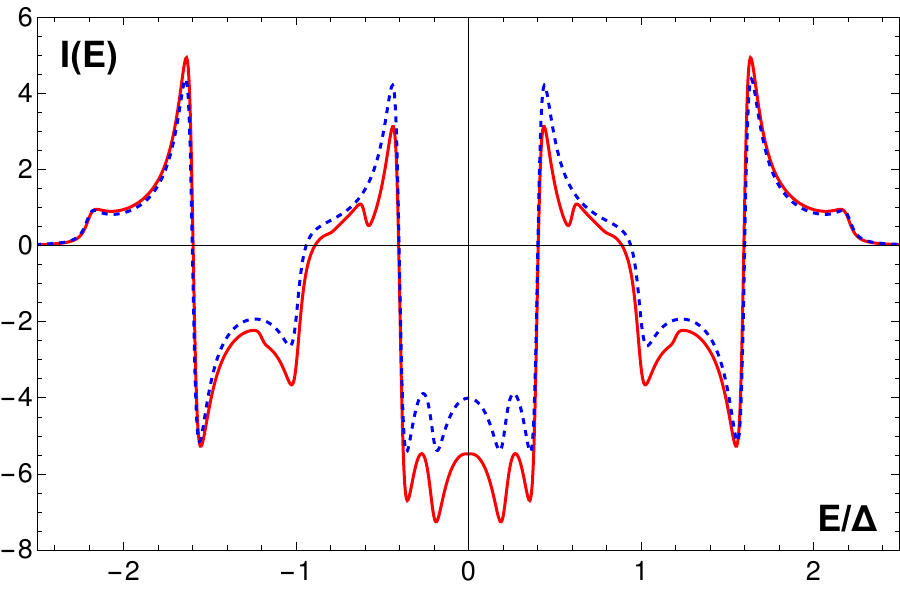}
\includegraphics[width=0.4\textwidth]{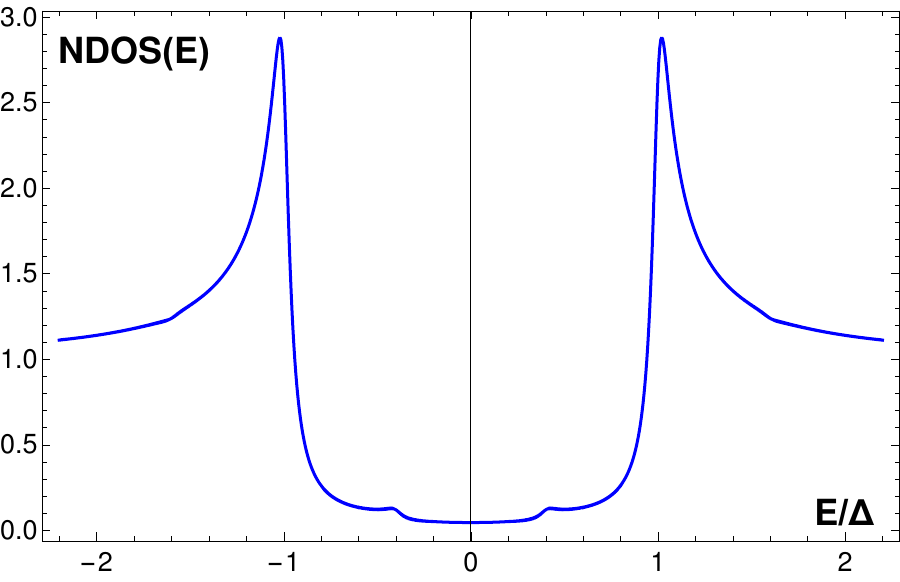}
\caption{Funnel TTJ, low transparency: (top) Spectral current (exact, thick; perturbative, dotted); (bottom) NDOS (exact and perturbative cannot be distinguished) ($V=0.6$, $\varphi_Q=\frac{2\pi}{3}$). Here $\Gamma_0=1,\Gamma_1=0.01, \Gamma_2=0.03$, $T=0.02$. Curves are vertically scaled by the squared transparency $T^2$.}
\label{fig:CurSpecPerturb}
\end{figure}

\begin{figure}[t]
\centering
\includegraphics[width=0.45\textwidth]{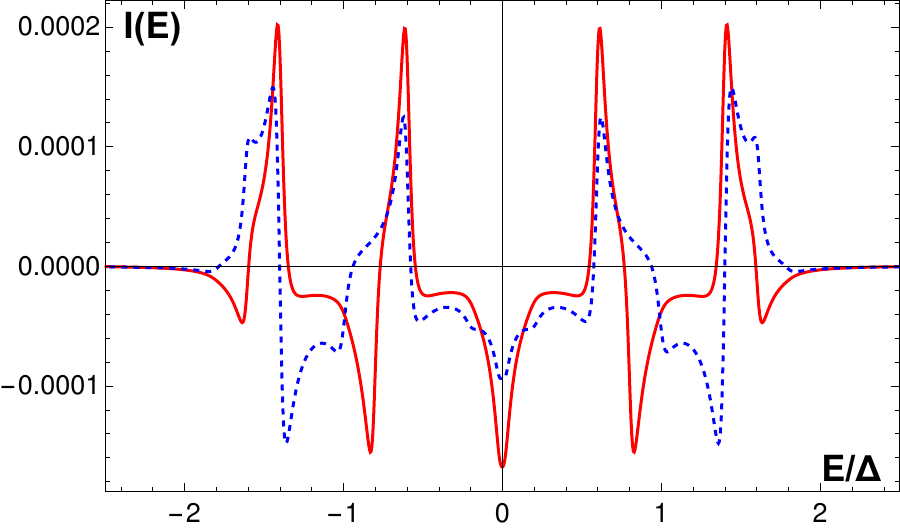}
\caption{Funnel TTJ, low transparency: Spectral current with moderate decoherence (exact, thick; perturbative, dotted) ($V=0.4$, $\varphi_Q=\frac{2\pi}{3}$, $\tau_d=0.4$, $\Gamma_0=1,\Gamma_1=\Gamma_2=0.002$, $T=0.001$).}
\label{fig:CurSpecPerturbDecoh}
\end{figure}

Let us now consider a "funnel" TTJ {\it out of equilibrium} (biased at voltages $V,-V$) with $\varepsilon\ll1$ and $T\ll1$ (see Section IV). Fig. \ref{fig:CurSpecPerturb} shows the exact result for the spectral current together with that of the perturbative calculation (see Sec.~\ref{sec:perturb}) in the coherent case ($\tau_d=0$), where the agreement between the two is remarkable. For moderate decoherence (Fig. \ref{fig:CurSpecPerturbDecoh}), the quality of the perturbative calculation degrades. Notice that to reproduce the exact current peaks of the numerical solution, a very small parameter $\varepsilon$ is required, because of resonances that are difficult to capture perturbatively. On the other hand, the NDOS is very close to that of the strongly coupled terminal $S_0$, i.e. a simple BCS spectrum. Indeed, the proximity effect in the node is here dominated by terminal $S_0$.

\begin{figure}[t]
\centering
\includegraphics[width=0.45\textwidth]{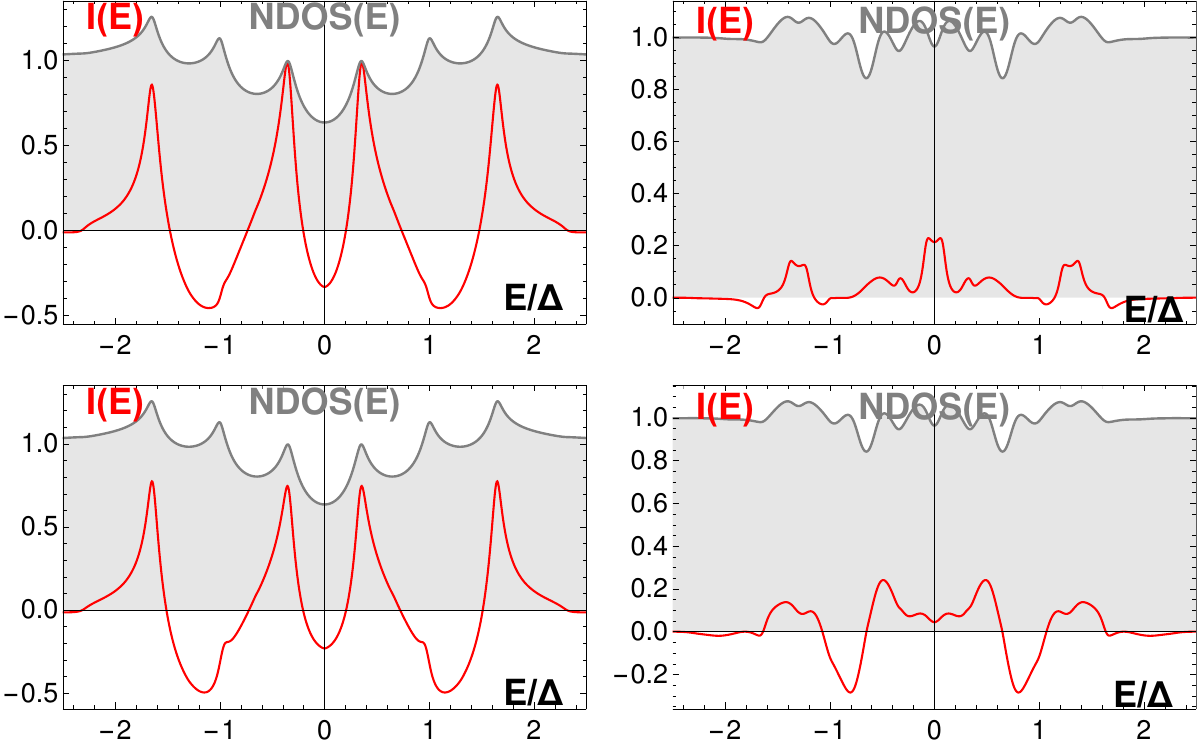}
\caption{Spectral current (red) and NDOS (grey) for a fully symmetric TTJ, $V=0.65$, $\varphi_Q=\frac{2\pi}{3}$, $\Gamma_0=\Gamma_1=\Gamma_2=1$ (left, $\tau_d=0.025$; right, $\tau_d=2$). Top : $T=0.1$. Bottom : $T=0.8$. Current for $\tau_d=2,T=0.1$ is scaled up by a factor $5$ to help visibility. The spectral current exhibits structures together with the NDOS, with marked sign changes.}
\label{fig:SpecCurTTJ}
\end{figure}

Let us now consider a fully symmetric TTJ,  $\gamma_1=\gamma_2=\gamma_0$. For low transparency $T\ll1$, the density of states displays sharp BCS-like structures centered at energy values corresponding to the six gap edges, e.g. given by $E=\pm \Delta, \pm(\Delta+V),\pm(\Delta-V)$.~\cite{Padurariu2} Due to higher order MAR processes, weaker structures also occur at $E=\pm(\Delta \pm nV)$ with $|n|>1$. Fig. \ref{fig:SpecCurTTJ} shows the spectral current in terminal $S_0$ (which reduces to the quartet contribution in this case), together with the NDOS, represented with a different scale. One sees that the spectral current peaks at the same energies as the NDOS. Remarkably, its sign changes from one structure to another. For large transparency (bottom panels of Fig. \ref{fig:SpecCurTTJ}), the peaks broaden but the spectral currents still oscillate from one structure to the next. For large decoherence and low transparency, one distinguishes three structures at $E \sim 0,\pm 2V$, of width $\sim \frac{1}{\tau_d}$, the latter being the mirror of the $E \sim 0$ region through Andreev reflection in terminals $1,2$. 

Finally, let us stress the consequences of these properties on the total current, which is obtained by the integration of the spectral current. As the spectral
current is an oscillating function of energy, which is very sensitive to the junction parameters that can shift or affect these oscillations, one can expect that the total current
can change sign when these parameters are varied. This will indeed be shown in the next sections.

\section{Phase and voltage dependence of the quartet currents.}

The quartet phase $\varphi_Q$ is stationary for any bias voltage $V$. Therefore the voltage appears as a novel control parameter for a DC Josephson current, in contrast with two-terminal junctions. As shown below and in next Section, the dependence of the quartet and the phase-MAR currents with $V$ is highly nontrivial and reflects the resonant dynamics of a driven quantum system. Here the quartet current $I_Q$ is calculated in terminal $S_0$  for a TTJ such that $\Gamma_1\sim\Gamma_2$, but with the ratio $\Gamma_1/\Gamma_0=\varepsilon$ taking any value.

\subsection{ Weakly asymmetric TTJ, $\varepsilon \simeq1$}

\begin{figure*}
\centering
\includegraphics[width=0.9\textwidth]{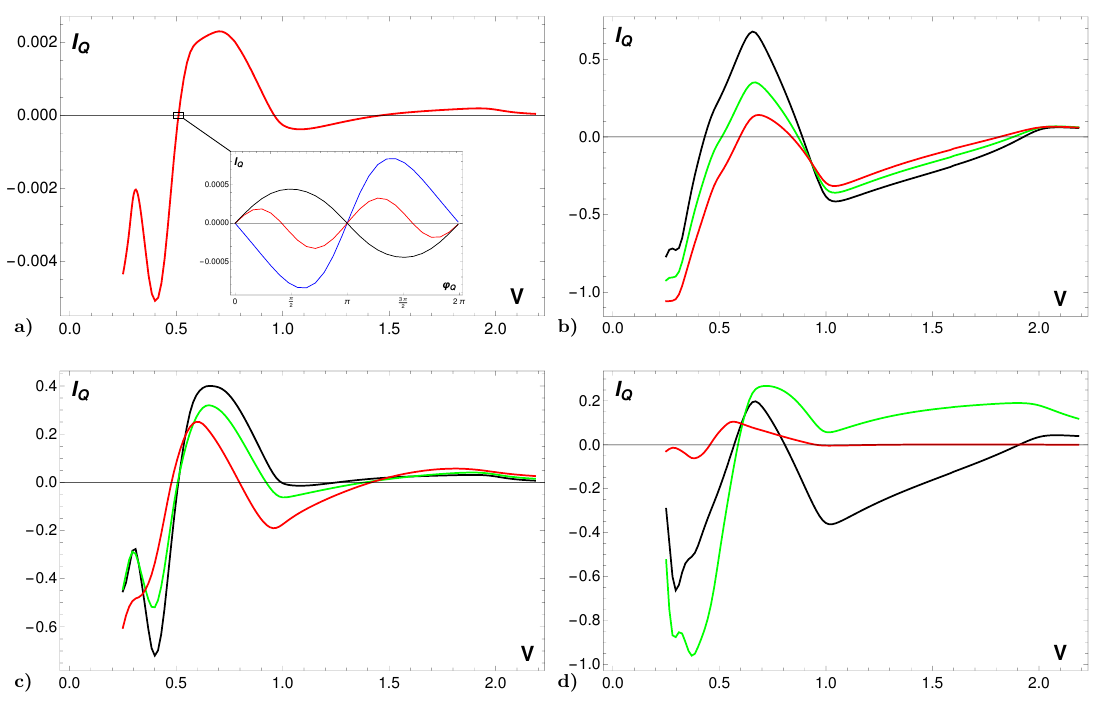}
\caption{(a) Weakly asymmetric TTJ quartet current as a function of $V$, low transparency ($\Gamma_0=1, \Gamma_1= \Gamma_2=1.5$, $T=0.01$, $\varphi_Q=\frac{2\pi}{3}$, $\tau_d=1.25$); (inset) corresponding quartet current-phase characteristics across a sign change. $V=0.495$ (blue); $0.5095$ (red, upscaled by factor $5$); $0.52$ (black); (b) TTJ quartet current as a function of $V$ for different transparencies, coherent case, $\varphi_Q=\frac{2\pi}{3}$. Curves are scaled (divided) by the transparency $T$ ($T=0.01$, black; $T=0.3$, green, $T=0.7$, red). $\varphi_Q=2\pi/3$, $\Gamma_0=1, \Gamma_1=1.5, \Gamma_2=0.6$, $\tau_d=0$; (c) TTJ quartet current as a function of $V$ for different transparencies, incoherent case; same scaling and parameters as (b), $\tau_d=1.25$; (d) TTJ quartet current as a function of $V$ for different decoherence rates, at low transparency ($\tau_d=0$, black; $\tau_d=0.3$, green, $\tau_d=3$, red). $\varphi_Q=2\pi/3$, $\Gamma_0=1, \Gamma_1=\Gamma_2=0.8$, $T=0.01$ (same scaling as (b)).} 
\label{fig:TTJsymetrique}
\end{figure*}

Let us first consider the situation close to the fully symmetric TTJ (see Fig.~\ref{fig:SchemaTTJ}a) with $\varepsilon \simeq1$.
We fix the quartet phase and plot the quartet DC current as a function of $V$, as an integral over energy of the spectral current. In all figures the current is in units of $\frac{e^2\Delta}{h}$. The behavior of a weakly asymmetric TTJ is rather complex, with several sign changes of the current (Fig. \ref{fig:TTJsymetrique}). Fig. \ref{fig:TTJsymetrique}b shows the effect of the transparency at zero decoherence, Fig. \ref{fig:TTJsymetrique}c corresponds to strong decoherence. In Fig. \ref{fig:TTJsymetrique}d transparency is fixed and decoherence is varied. The calculation is limited by numerics for small $V$ and other sign changes may occur in the very low $V$ regime (where the current experiences strong oscillations with $V$). A similar behavior was obtained in a separate quantum dot model calculation.~\cite{Regis1,Regis2,Regis3}

 Figure \ref{fig:TTJsymetrique}a shows the quartet current-phase characteristics close to a sign change, around $V=0.5\Delta$. The anharmonicity is strong very close to $V=0.5\Delta$ at the "$\pi-0$" transition. A general trend is as follows: "$\pi$"-junction  at low voltage, "$0$"-junction at intermediate voltage and "$\pi$"-junction at higher voltage. 
 
The quartet current  vanishes at high voltages (Figs. \ref{fig:TTJsymetrique}b, \ref{fig:TTJsymetrique}c, and \ref{fig:TTJsymetrique}d) regardless of transparency and decoherence  ($V>2\Delta$ signals the quasiparticle transfer onset between the side and central leads).
 
Changing the transparency does not affect ``much'' the overall shape of the quartet current - voltage characteristics: the voltage at which the current changes sign then depends on the transparency $T$, especially in the presence of decoherence (Figs. \ref{fig:TTJsymetrique}b and \ref{fig:TTJsymetrique}c), and modifies its amplitude (we stress that the curves have been rescaled by the transparency for convenience). The introduction of moderate decoherence $\tau_d=1$  reduces somehow the amplitude of the quartet current. Fig. \ref{fig:TTJsymetrique}d shows the effect of decoherence in the tunnel regime, where we see that strong decoherence cancels the sign change at large voltages.

Let us emphasize that the occurrence of a new control parameter $V$ for the quartet current offers the following possibility : by changing the overall transparency of the TTJ, or that of one or two contacts only, one can change the sign of the current, provided the voltage $V$ is suitably chosen. This extends the range of phenomena causing sign changes in a Josephson current. This might be useful in devices where commuting a junction from "$\pi$" to "$0$" is required.

\subsection{"Funnel" TTJ}

\begin{figure*}
\centering
\includegraphics[width=0.9\textwidth]{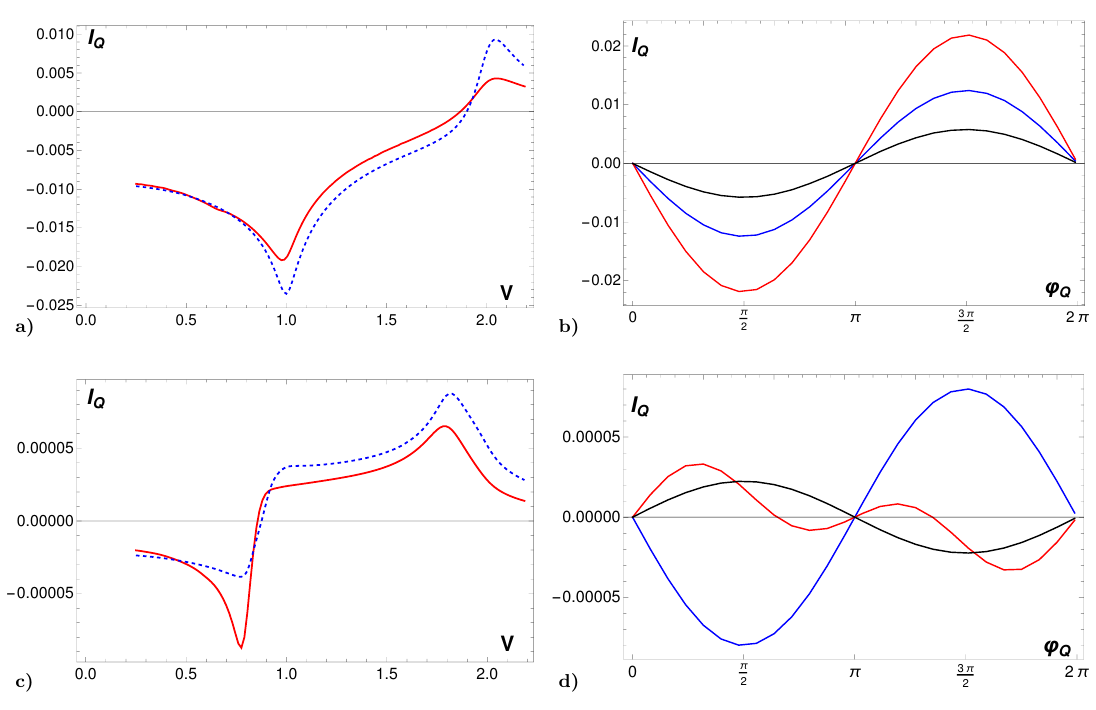}
\caption{(a) Funnel TTJ quartet current as a function of voltage, coherent case (exact, thick; perturbative, dotted), $\varphi_Q=2\pi/3$, $\tau_d=0$; (b) Quartet current-phase characteristics, $V=0.5$ (blue), $1$ (red), $1.5$ (black); $\Gamma_0=1, \Gamma_1= \Gamma_2=0.03$, $T=1$, $\varphi_Q=2\pi/3$, $\tau_d=0$;  (c) Funnel TTJ quartet current as a function of 
voltage (exact, thick; perturbative, thin), $ \varphi_Q=2\pi/3$, 
$\tau_d=0.4$; (d) Quartet current-phase characteristics, $V=0.8$ (blue), $0.855$ (red, vertical upscaling by a factor $500$), $0.9$ (black); $\Gamma_0=1, \Gamma_1= \Gamma_2=0.005$, $T=0.1$, $ \varphi_Q=2\pi/3$, 
$\tau_d=0.4$.}
\label{fig:CurPerturb}
\end{figure*}

Let us now consider a "funnel" TTJ with $\varepsilon \ll 1$ (Fig. \ref{fig:SchemaTTJ}b), and compare the exact current with that found by second order expansion in $\varepsilon$ (see Sec.~\ref{sec:perturb}) The result is displayed in Fig. \ref{fig:CurPerturb}a. In the coherent case, the agreement with the perturbative result is qualitatively good if $\varepsilon$ is small enough. Notice that the total current is the result of an integral over a strongly oscillating function, and the result is therefore quite sensitive to approximations.
Remarkably, the sign of the quartet current is always negative, apart for high values of $V$. This corresponds to a "$\pi$" quartet junction. 
Fig. \ref{fig:CurPerturb}b shows the quartet current-phase characteristics for a given $V$, which is close to a harmonic relation $I_Q=I_{Qc}\sin\varphi_Q$, with $I_{Qc}<0$. The quartet critical current peaks around $V=\Delta$. 

Conversely, with moderate decoherence, the agreement with the perturbative treatment in $\varepsilon$ is less good, especially for intermediate voltages (Fig. \ref{fig:CurPerturb}c): in this region, the quartet current changes sign, from a "$\pi$"-junction (low voltage) to a "$0$"-junction (high voltage). The quartet characteristics $I_Q(\varphi_Q)$ at fixed $V$ becomes strongly anharmonic in the region of the sign change (Fig. \ref{fig:CurPerturb}d). 

\begin{figure}[t]
\centering
\includegraphics[width=0.45\textwidth]{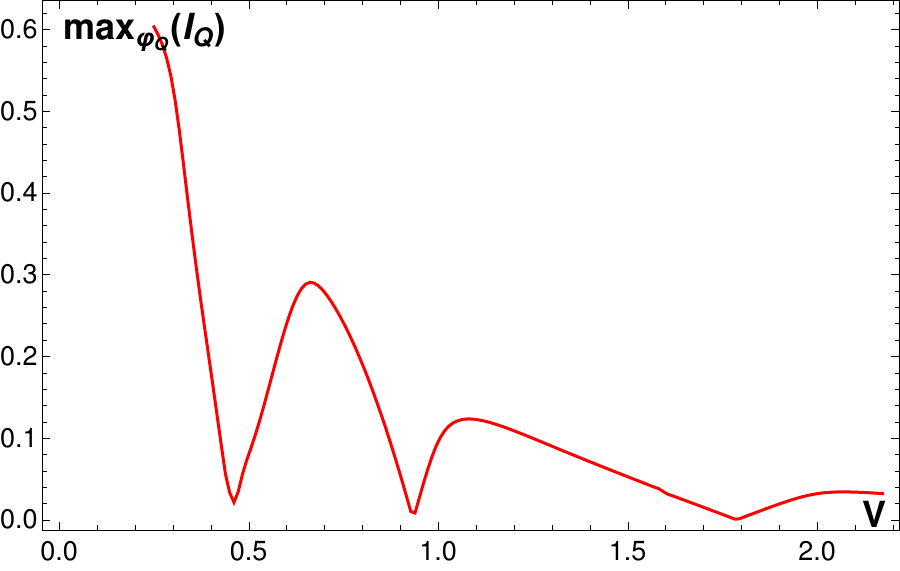}
\caption{Nonmonotonous variation with $V$ of the TTJ maximum quartet current (red), taken on the quartet phase ($\Gamma_0=\Gamma_1= \Gamma_2=1$, $T=1$, $\tau_d=0$).} 
\label{fig:MaxQuartetCurrent}
\end{figure}

Let us end this Section by proposing an indirect experimental detection of the sign changes, when a phase-sensitive experiment is not possible. Fig. \ref{fig:MaxQuartetCurrent} shows the maximum ("critical") quartet current, taken on all possible quartet phase values. It displays sharp minima when the phase-sensitive current changes sign. Similar trends were observed in ferromagnetic Josephson junctions as a witness of $0/\pi$ transitions, as a function of the junction length.~\cite{SFS} Here, such minima observed as a function of $V$ are hardly explainable by conventional models based on extrinsic synchronization by the outer circuit, as often advocated in experiments against the mesoscopic quartet mechanism.

\section{MAR currents and the balance with quartet currents}
\subsection{Phase symmetry of MAR currents}
We first consider a junction which is symmetric in terminals $1,2$, i.e. $\Gamma_1 = \Gamma_2$. The full MAR current is defined as $I_{MAR}=\frac{e^2}{2h}T\Gamma_1(I^1-I^2)$, and there is no MAR current in terminal $S_0$. The phase-dependent part of the MAR current $I_{phMAR}^i$ is defined by subtracting from the full MAR current its average on the quartet phase $\varphi_Q$:
\begin{equation}
I_{phMAR}^i=I_{MAR}^i- \langle I_{MAR}^i \rangle_{\varphi_Q}
\end{equation}

\begin{figure}
\centering
\includegraphics[width=0.45\textwidth]{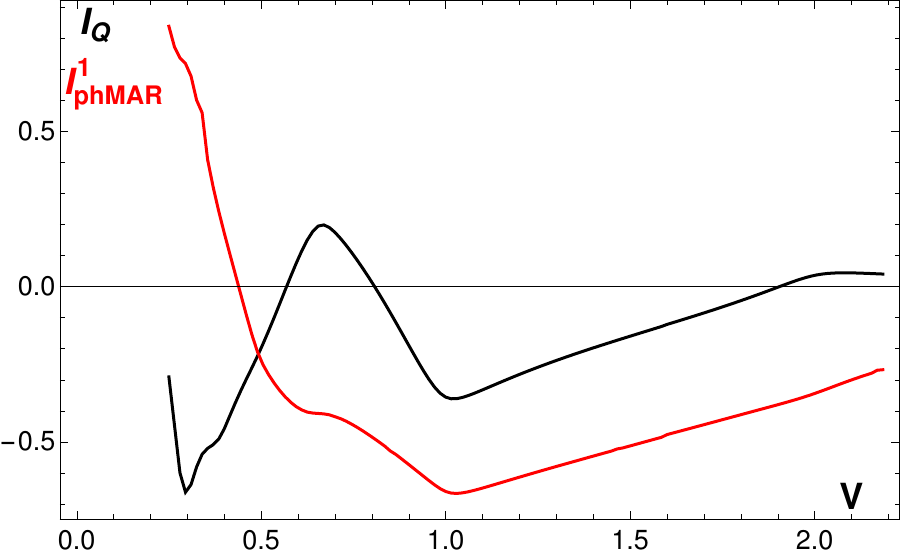}
\includegraphics[width=0.45\textwidth]{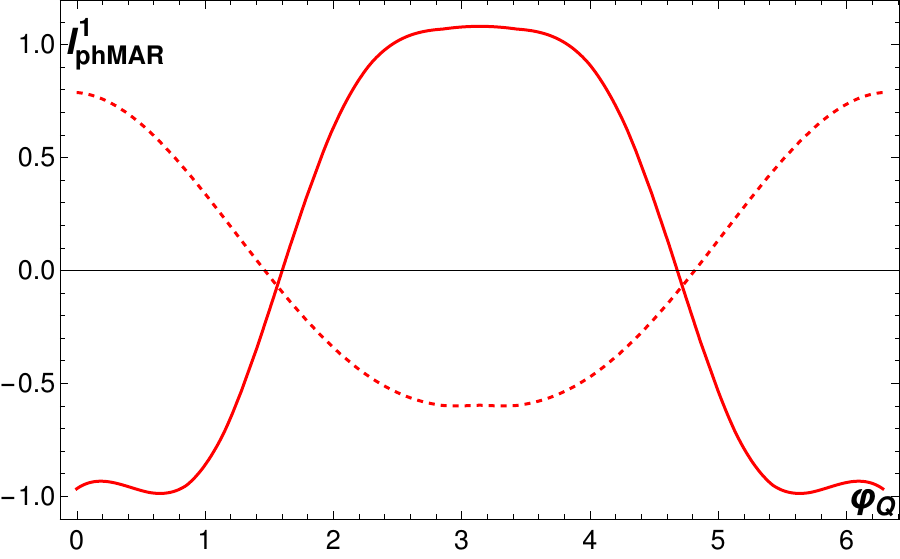}
\caption{(top) TTJ quartet current (black) and phase MAR current (red) as a function of $V$ ($\Gamma_0=1, \Gamma_1= \Gamma_2=0.8$, $\varphi_Q=2\pi/3$, $T=0.01$, $\tau_d=0$); (bottom) Phase MAR current-phase characteristics across a sign change. $V=0.25$ (full); $0.76$ (dotted).} 
\label{fig:ComparisonQphmarSym}
\end{figure}

The phase-MAR current is represented as a function of the phase $\varphi_Q$ on Fig. \ref{fig:ComparisonQphmarSym}. It is an even function of $\varphi_Q$, contrarily to the quartet current which is odd.  As a function of voltage, the phase-dependent MAR current also presents sign changes, at values of $V$ different from those making the quartet component change sign.

\subsection{A TTJ is a $\varphi$-junction}
Owing to the different symmetries of quartet and phase-MAR currents with respect to the quartet phase, their superposition yields a phase shift, that makes a TTJ a tunable "$\varphi$-junction": the phase shift indeed depends on the parameters of the TTJ : bias voltage $V$, transparency, decoherence, and also on the two relevant asymmetries: that comparing contact $0$ to contacts $(1,2)$ (see the "funnel" case), and that mutually comparing contacts $1,2$.
\begin{figure}
\centering
\includegraphics[width=0.45\textwidth]{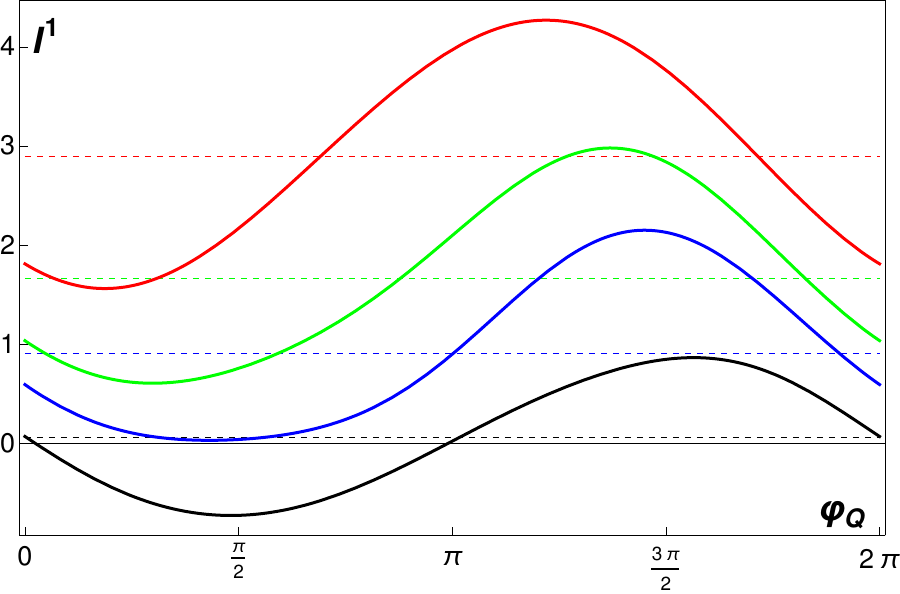}
\includegraphics[width=0.45\textwidth]{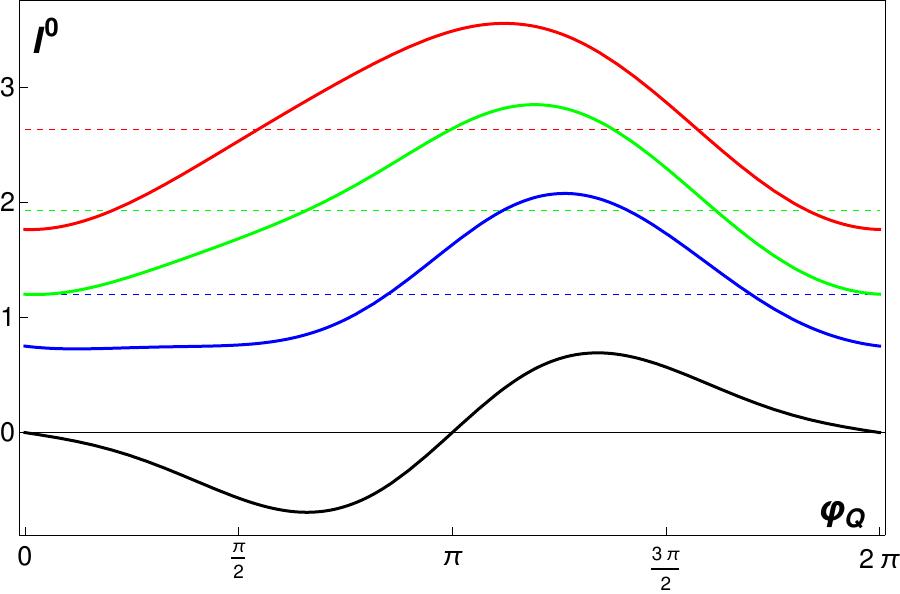}
\caption{(top) Current in $S_1$ (normalized by a factor $\gamma_1$) showing different phase shifts, with $\Gamma_0=1, \Gamma_1= \Gamma_2=0.2$, $\tau_d=0.5$, $V=0.2$ (black, quartet-dominated current), $0.45$ (blue), $0.55$ (green), $0.7$ (red, MAR-dominated current)). Dotted lines feature the phase-independent MAR current; (bottom) current in $S_0$,
showing different phase shifts, with $V=0.4$, $\tau_d=1$, $\Gamma_0=1$, and $\Gamma_1= \Gamma_2=1$ (black, quartet-dominated current), $\Gamma_1=1.2, \Gamma_2=0.8$ (blue), $\Gamma_1=1.4,  \Gamma_2=0.7$ (green), $\Gamma_1=2,  \Gamma_2=0.5$ (red, MAR-dominated current)).} 
\label{fig:thetaV}
\end{figure}
\begin{figure}
\centering
\includegraphics[width=0.45\textwidth]{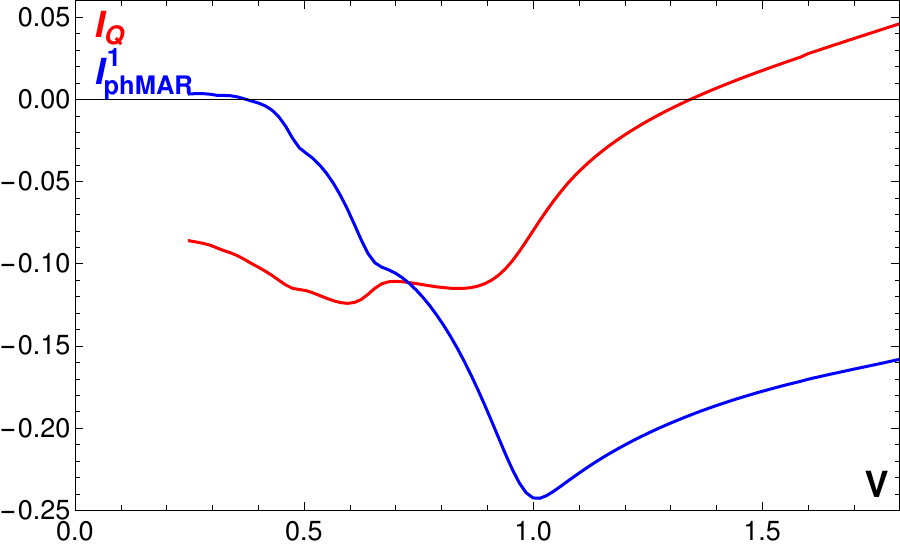}
\includegraphics[width=0.45\textwidth]{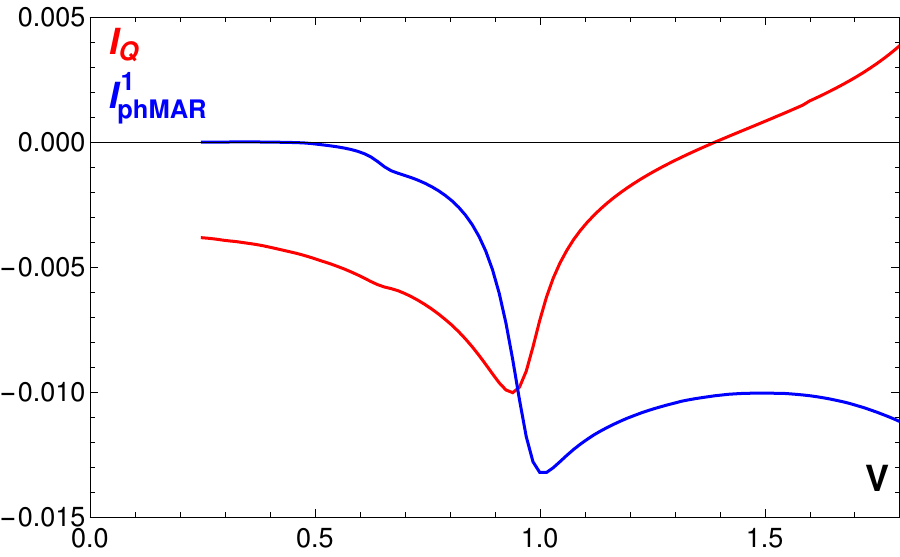}
\caption{TTJ quartet current (red) and phase MAR current (blue) as a function of $V$ ($\Gamma_0=1, \Gamma_1= \Gamma_2=\varepsilon \Gamma_0$, $\varphi_Q=2\pi/3$, $\tau_d=0$); (top) $\varepsilon=0.1$; (bottom) $\varepsilon=0.02$} 
\label{fig:ComparisonQphmarFunnel}
\end{figure}
\begin{figure}
\centering
\includegraphics[width=0.45\textwidth]{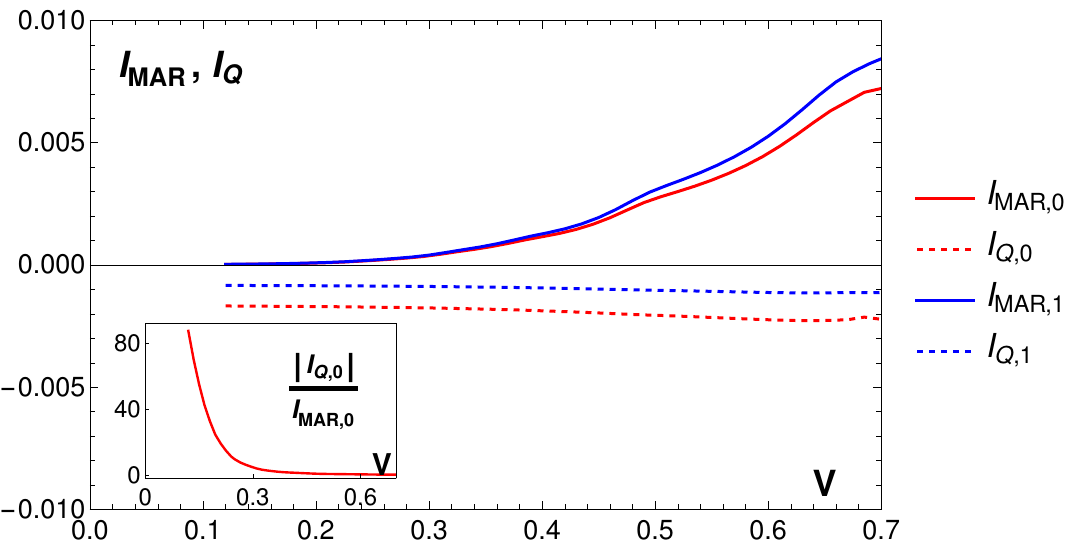}
\caption{Quartet and full MAR currents in a fully asymmetric TTJ, $\Gamma_0=1,\Gamma_1=0.2,\Gamma_1=0.05$. MAR currents in terminals $S_0$ and $S_1$ strongly decay at low voltages, contrarily to quartet currents. Current sign convention is inverted in terminals $1,2$ to help comparison. Notice that here $I_{q0}\simeq2I_{q1}\simeq2I_{q2}$.} 
\label{fig:asymMAR}
\end{figure}

This is illustrated by two examples: first, within $(S_1,S_2)$ symmetry, the current in terminal $S_1$ superimposes quartet and MAR components thus displays a phase shift, which in Figure \ref{fig:thetaV}(top) varies with the voltage. As $V$ increases, it passes from a quartet-dominated current, nearly zero at zero phase, to a MAR-dominated current, with phase-MAR component maximal near zero phase. Second, with an $(S_1,S_2)$ asymmetry, the current in terminal $S_0$ contains MARs in addition to quartets and displays a phase shift, here at fixed voltage and varying the asymmetry (Figure \ref{fig:thetaV}(bottom)). It goes from a quartet regime to a MAR regime as asymmetry is increased. Provided phase-MARs are not negligible, such a phase shift could be observed in a phase-sensitive experiment as proposed in Ref. \onlinecite{QuartetSQUID}.

\subsection{Conditions to observe quartets against MARs}
In a non phase-sensitive transport experiment, and in the case where MARs are observed in the full $(V_1, V_2)$ conductance map, care must be taken in interpreting the $V-I$ anomaly observed on the "quartet line" $V_2=-V_1$. For a rather symmetric TTJ one finds that $I_Q$ and $I_{phMAR}$ have comparable magnitudes, even at low voltage (Fig. \ref{fig:ComparisonQphmarSym}).This is due to the fact that in the present model simulating a cavity with a dense level spectrum, a symmetric TTJ behaves as a resonant junction and therefore MARs can be strong even at low voltage. On the contrary, when $\Gamma_1,\Gamma_2\ll\Gamma_0$, the phase-MAR current becomes very small at low voltages and is dominated by the quartet current (Fig. \ref{fig:ComparisonQphmarFunnel}). 

As shown above, a funnel asymmetry is sufficient to make phase-MAR very small. This is because those processes involve coherently all three contacts and taking $\varepsilon\ll1$ suppresses multiple Andreev reflections between $S_0$ and both ($S_1,S_2$). It also trivially hampers two-terminal MARs occurring either between $S_0$ and $S_1$, or between $S_0$ and $S_2$, which participate to the phase-independent MAR current. Nevertheless, if couplings between the node and the leads $S_1,S_2$ are comparable albeit small, MARs between those contacts (with voltage difference $2V$)  become resonant and can overcome quartets. Fig. \ref{fig:asymMAR} shows that if one instead takes $\Gamma_0\gg\Gamma_1\gg \Gamma_2$ (or $\Gamma_0\gg \Gamma_2\gg\Gamma_1$) (full asymmetry, see Fig. \ref{fig:SchemaTTJ}c), all two-terminal MARs are suppressed at low $V$ and the total MAR current becomes negligible compared to the quartet current below $V=0.2\Delta$ in the coherent case.

\section{Discussion and conclusion}

In this work we have studied in detail the voltage and phase dependence of the two DC current components involving quartets in a three-terminal Josephson junction biased with opposite voltages : quartet current and phase-MAR current. The salient feature is the sign changes occurring as the voltage is varied. The quartet current switches from a $\pi$ to a $0$ junction.
These sign changes are absent at low voltage when the junction is strongly asymmetric in terms of the coupling to the unbiased "source" terminal compared to the two others ("funnel junction"). This, and the agreement between a perturbative and the exact calculation, points towards non-adiabatic effects as being responsible for such sign changes. 
This is similar to the behavior reported in quantum dot junctions.~\cite{Harvard,Regis1,Regis2,Regis3} In the funnel junction, a wide pseudogap in the normal density of states protects from nonadiabatic transitions between negative and positive energy Andreev bound states, and the finite voltage behavior emerges adiabatically from the zero-voltage one which manifests a $\pi$-junction.  

The effect of decoherence has been also investigated, as well as the overall transparency of the contacts. This leads to shifts of the sign changes, that sensitively depend on the multiple Andreev reflections occurring at the contacts. At low voltages, phase-MAR currents become negligible compared to quartet currents in a funnel asymmetric TTJ. All MAR currents become negligible if the low-voltage TTJ is doubly asymmetric. When phase-MARs are not negligible, which is the generic situation, the TTJ becomes a controllable $\varphi$-junction.

Probing the sign changes of the quartet current, more generally the $\varphi$-TTJ, requires a full quartet SQUID geometry.~\cite{QuartetSQUID} A less demanding probe is the strongly non-monotonic variation of the quartet "critical" current $I_{qc}$, that can be observed in a direct transport experiment, as measured by the width of the Josephson-like anomaly found on the quartet line. This non-monotonic dependence and the strong sensitivity of the positions of the $I_{Qc}$ minima with voltage are a hallmark of quartet processes. They are directly related to the mesoscopic nature of the quartet process and can thus be distinguished from other synchronization mechanisms due to an external circuit impedance, stronger than that of the junction itself : such an impedance has no reason to depend sensitively on the details (transparency, decoherence, asymmetries) of the TTJ.

Potential extensions of this work include the computation of the noise characteristics of the TTJ within the circuit theory framework. In previous works where the diffusive node between the superconducting leads is replaced by a system of quantum dots~\cite{Regis1,Jacquet} showed that at low temperature, the Fano factors relevant to the TTJ vanish for decreasing voltage, as expected for a Josephson-like signal (the same should be true in the context of circuit theory).   

\begin{acknowledgments}
This work received support from the French government under the France 2030 investment plan, as part of the Initiative d'Excellence d'Aix-Marseille Universit\'e - A*MIDEX (authors J. R., T. J., L.R., and T. M.). Centre de Calcul Intensif d'Aix-Marseille is acknowledged for granting access to its high performance computing resources.
\end{acknowledgments}

\end{document}